# A Critical Reexamination of the Electrostatic Aharonov-Bohm Effect

**Allan Walstad**
Department of Physics, University of Pittsburgh at Johnstown, Johnstown, PA 15904 USA
e-mail: awalstad@pitt.edu   phone: 814-269-2974   fax: 814-269-2022

**Abstract**  This paper undertakes a critical reexamination of the electrostatic version of the Aharonov-Bohm ("AB") effect.  The conclusions are as follows:  1. Aharonov and Bohm's 1959 exposition is invalid because it does not consider the wavefunction of the entire system, including the source of electrostatic potential.  2. As originally proposed, the electrostatic AB effect does not exist.  Perhaps surprisingly, this conclusion holds despite the relativistic covariance of the electromagnetic four-potential combined with the well-established magnetic AB effect.  3. Although the authors attempted, in a 1961 paper, to demonstrate that consideration of the entire system would not change their result, they inadvertently assumed the desired outcome in their analysis.  4. Claimed observations of the electrostatic AB or an analogue thereof are shown to be mistaken.

**Keywords**  Aharonov-Bohm effect

## 1  Introduction

The Aharonov-Bohm ("AB") effect refers to charged-particle quantum interference phenomena that can only be ascribed to the action of electromagnetic potentials, as the particles themselves propagate entirely through electric and magnetic field-free regions. Two versions were proposed in 1959 by Aharonov and Bohm [1], one mediated exclusively by the electrostatic potential (the electrostatic effect), the other by the vector potential (the magnetic effect).  The latter was soon observed experimentally [2,3], but confirmation of the former proved more elusive.  As this paper demonstrates, the proposed electrostatic AB effect rests on an elementary theoretical error.  An oft-cited observation claim [6] is also shown to be mistaken.

Following Aharonov and Bohm (with modified notation), we recall that the Hamiltonian for a particle of charge $q$ in a region of electrostatic potential $V$ can be written $H = H_o + qV$, where $H_o$ is the Hamiltonian in the absence of $V$.  Schrodinger's equation for the Hamiltonian H is satisfied by

$$\psi = \psi_o\, e^{i\varphi}, \qquad \varphi = (q/\hbar) \int V(t)\, dt \qquad (1)$$

where $\psi_o$ satisfies Schrodinger's equation for $H_o$.  The effect of $V$ is thus to introduce a phase $\varphi$ into the wavefunction.

If the wavepacket of a particle is split and allowed to propagate to a common destination via two different paths, and if $V(t)$ is different on those two paths, then it seems the difference in phase should result in observable interference effects.  Aharonov and Bohm described an idealized experiment in which the wavepackets travel through long, narrow metal cylinders, with $V$ being turned on and off again for one cylinder only, while the wavepackets are within the apparatus (Fig. 1). They suggested that by changing the magnitude and duration of $V$, the resulting interference pattern would be shifted. Yet, the interiors of the cylinders are, to high accuracy, field-free regions. This proposed ability of an electrostatic potential to cause a physical effect, in the absence of an electric field or force acting on the particle, was counter-intuitive for many physicists at the time.

Given that the electrostatic and vector potentials $V$ and $\mathbf{A}$ are Lorentz frame-dependent components of a relativistic four-vector, Aharonov and Bohm noted that the expression for electromagnetic phase must be generalized as follows:

$$\varphi = (q/\hbar) \left[ \int V \, dt - \int \mathbf{A} \cdot d\mathbf{r} \right]. \tag{2}$$

Outside a long, thin solenoid, there is an azimuthal vector potential but no appreciable magnetic field. Quantum interference between trajectories passing on opposite sides of the solenoid will depend on the phase shift $(q/\hbar) \int \mathbf{A} \cdot d\mathbf{r} = (q/\hbar) \, \Phi$, where $\Phi$ is the magnetic flux through the solenoid. That is the magnetic AB effect.

## 2 The Problem

As Sakurai indicates [4], the proposed electrostatic AB effect may be understood as an example of the energy-frequency relation $E = h\nu$, according to which the rate of phase generated depends on the energy. If we can arrange for the electrostatic potential energy $qV$ of a charged particle to differ along different interfering paths, without a compensating change in kinetic energy, then the phase difference and resulting interference pattern should be affected. These conditions appear to be met in the idealized experiment of Fig. 1.

But this interpretation raises a problem. The charged particle is one component of a larger system that includes the source of electrostatic potential acting on it. The wavefunction of this system evolves in a many-dimensional configuration space. Quantum interference occurs between different paths of the entire system ending at a given point in that configuration space. Conservation of energy requires that a change in energy of the particle be accompanied by an equal and opposite change in energy of the rest of the system. If the particle's energy changes by $\Delta E$ for a time $t$, the extra phase for the particle is $\Delta E \, t/\hbar$. The energy of the rest of the system changes by $-\Delta E$ for the same time $t$, giving a phase of $-\Delta E \, t/\hbar$. The overall phase is unchanged and does not depend on the path taken by the charged particle.

Just how is the energy of the rest of the system affected by the presence or absence of $q$? In order to change the voltage of a cylinder, charge is brought to its surface from elsewhere. Even if no net force acts on $q$, nevertheless $q$ exerts forces on the approaching charges, thereby changing the amount of work required to charge the cylinder. This work comes at the expense of an energy source which must be considered part of the overall system for which quantum interference occurs.

A correct analysis of interference effects starts with recognizing that phase is generated not just by energy but by the Lagrangian along any given path in configuration space:

$$\varphi = (1/\hbar) \int L \, dt = (1/\hbar) \int (E \, dt - p_i \, dq_i) \qquad (3)$$

where the $q_i$ and $p_i$ are the coordinates and conjugate momenta characterizing the entire system. Since energy is conserved, and since the duration of the experiment is the same for all components of the system as observed in any given Lorentz frame, the integral of $E \, dt$ is necessarily the same for all accessible paths. Since the proposed electrostatic AB effect depends on a difference in the energy integral for different paths, *it does not exist*.

By contrast, different parts of the system do not necessarily undergo identical spatial displacements. Thus, although total momentum is conserved, the integral of $p_i \, dq_i$ can differ for different accessible paths in configuration space. In an experiment intended to detect an electrostatic AB effect, it is possible that the electric forces exerted by the charged particle on different parts of the apparatus might thereby result in an observable phase shift (while the reaction forces on the charged particle itself cancel out). Nevertheless, *being mediated by electric forces*, such a phase shift would not be an AB effect.

In the magnetic version of the AB effect the phase gradient $(q/\hbar)\,\mathbf{A}$ is to be understood in terms of an electromagnetic momentum acquired by a charged particle in the presence of a vector potential. As the particle travels past the solenoid, the electromagnetic momentum acting through the particle's displacement generates a phase which depends on the path. Conservation of momentum requires, by whatever mechanism, that the solenoid acquire an equal and opposite momentum. Nevertheless, because the solenoid does not undergo an appreciable displacement, there is no corresponding phase generated, leaving us with only the phase difference due to the motion of the charged particle. Thus, the existence of the experimentally well-established magnetic AB effect is entirely consistent with the non-existence of an electrostatic AB effect, despite the mixing of components of the electromagnetic four-potential by Lorentz transformation among different frames, which one might have expected would inextricably entangle the two mechanisms.

### 3. Aharonov and Bohm's 1961 Paper and the Wavefunction of the Entire System

In the abstract to a 1961 paper [5], Aharonov and Bohm stated, "We…extend our treatment to include the sources of potentials quantum-mechanically, and we show that when this is done, the same results are obtained as those of our first paper…." For our purposes, the relevant part is their section 3. There, the authors wish to demonstrate that with suitable simplifying assumptions the overall wavefunction factors into a wavefunction for the source (having no dependence on the position of the charged particle) times a single-particle wavefunction obeying Schrodinger's equation with the time-dependent electrostatic potential $V$. If that were the case, then the phase difference would be determined solely by the single-particle wavefunction, and the result of the 1959 paper would be confirmed.

Their Equation 11 is Schrodinger's equation for the entire system,

$$i\hbar \frac{\partial}{\partial t}\Psi(\mathbf{x},...y_i...,t) = [H_e + H_S + V(\mathbf{x},...y_i...)]\Psi(\mathbf{x},...y_i...,t) \qquad (11AB)$$

with **x** the location of the particle and $y_i$ the coordinates characterizing the parts of the apparatus (the source of potential). We see that the Hamiltonian has been broken into three parts: one each for the particle and the apparatus, and an interaction potential energy $V$. (Note: Elsewhere in this paper $V$ is used to denote electric potential, not potential energy.) They seek to treat the apparatus via a WKB approximation, writing its wavefunction (in their Equation 13) in terms of a magnitude $R$ and phase $S/\hbar$ as

$$\phi(...y_i...) = R(...y_i...,t)e^{iS(...y_i...,t)/\hbar}. \qquad (13AB)$$

$S$ and $R$ are said to obey, respectively, their Equations 14 and 16:

$$\frac{\partial S}{\partial t} + \sum_i \frac{1}{2M_i}\left(\frac{\partial S}{\partial y_i}\right)^2 + W(...y_i...) = 0 \qquad (14AB)$$

$$\frac{\partial P}{\partial t} + \sum_i \frac{\partial}{\partial y_i}\left(\frac{p_i}{M_i}P\right) = 0 \qquad (16AB)$$

where in (16AB) the substitution $P = R^2$ was made. [In (14AB) an obvious typographical error in the original paper has been corrected by replacing $V(\ldots y_i\ldots)$ with $W(\ldots y_i\ldots)$, the potential energy of interaction of all parts of the source with each other, introduced in Equation 12 of the original paper.]

The problem with (13AB), (14AB), and (16AB) is that they do not allow any possibility for the charged particle to affect the wavefunction of the apparatus, as the position of the particle does not appear in those equations. Aharonov and Bohm have in effect already assumed that because the apparatus is massive, it goes about its way as though the charged particle were absent; thereby, they have already assumed that the overall wavefunction factors as desired. It does not. The change in motion of the (parts of the) apparatus due to forces exerted by the particle may indeed be tiny, but only a tiny change in motion is required in order to effect a phase change of order unity, sufficient to invalidate their conclusion.

In the same paper, Aharonov and Bohm appear to acknowledge that the idealized experiment from their 1959 paper (discussed earlier in connection with Fig. 1) does not actually satisfy the conditions for quantum interference. One cylinder is raised to voltage $V$ for a time and then grounded again. All this must take place while the particle is within a cylinder, to avoid the particle's being acted on by electric fields on the way into or out of the apparatus. To achieve a given voltage requires that a different amount of charge be placed on the cylinder if the particle is present within it than if not. (If we imagine spheres instead of cylinders, it is easy to see that the difference will be equal in magnitude to *q* of the particle.) So the net amount of charge drawn from the voltage source and conducted to ground constitutes a record of the path taken by the particle, and interference cannot occur.

The authors therefore offered additional, highly idealized gedanken experiments that would overcome the above objection by not allowing for information to be registered about which path the particle had taken. It is not necessary to address these scenarios here. It is sufficient to point out that any resulting interference effect would necessarily be mediated by electric forces acting on the components of the apparatus, as noted earlier, and would thus not fulfill the definition of an AB effect.

## 4  Electrostatic AB Effect Observed?

Van Oudenaarden et al. [6] observed oscillations in the conductance of a mesoscopic metal ring as a function both of the applied voltage and of an externally-imposed magnetic flux through the ring. They interpreted the results as being due to a combined electrostatic and magnetic Aharonov-Bohm effect. On the contrary, as explained below, what the authors ascribed to an electrostatic AB effect is simply the result of electrostatic forces acting directly on charged particles.

As illustrated in Fig. 2 (adapted from [6]), the metal ring is interrupted on opposite sides by tunnel junctions. Current flow from source to drain in response to an applied voltage *V* requires that electrons tunnel across these junctions. Tunneling results in an electron and a hole propagating in opposite directions from the junction. If the electron and hole propagate to the drain and source, respectively, they contribute to the current. If, however, the electron and hole travel around the ring and recombine at the other junction, they do not contribute to the current. Although the ring is not superconducting, the temperature is low enough that phase coherence is maintained despite the diffusive motion of the electron and hole. Quantum interference can therefore affect the likelihood that a tunneling event will contribute to the current, and thereby it can affect the conductance of the ring.

Consider the following two paths of the system between the same starting and ending state:

a) There is no tunneling event.
b) There is a tunneling event at one junction, after which the electron and hole travel in opposite directions around the ring and recombine at the other junction.

If the phase difference between these paths is $2\pi$ or a whole multiple thereof, interference will reinforce a result that does not contribute to the current, and conductance will be suppressed. If the phase difference is $\pi$ or an odd multiple thereof, then this result will suffer cancellation and conductance will be enhanced. At constant *V*, conductance maxima are observed to occur at magnetic flux values differing by $\Delta\Phi = h/e$ (where *e* is the charge quantum), which is unambiguously what we expect from the magnetic Aharonov-Bohm effect.

The oscillations of conductance with magnetic flux are modulated periodically as a function of *V*. This dependence on *V* is interpreted by [6] in terms of an electrostatic AB effect. On this interpretation, a phase $eV t_o / \hbar$ is generated during the time $t_o$ between tunneling and recombination, while the electron and hole traverse their respective sides of the ring at a potential difference $V$. Setting $e\,\Delta V\,t_o/\hbar = 2\pi$, where $\Delta V$ is the change in $V$ between conductance peaks, we can infer $t_o$.

The relationship between $t_o$ and $\Delta V$ is correct, but it could not be an example of an electrostatic AB effect. It is the defining premise of an AB effect that interfering particles travel through field-free regions, but in the experiment of [6] the electron and hole are subject to an electric field in tunneling across a junction. It is easily seen that the momentum thereby acquired is responsible for the interference, as follows.

Let $s$ be the total distance traveled by an electron or hole in its diffusive motion from one junction to the other. At speed $v$ the time required is $t_o = s/v$ and the deBroglie wavelength is $\lambda = h/mv$. The phase generated by mechanical momentum is $\varphi = 2\pi s/\lambda = 2\pi s m v/h$. A change $\Delta V$ in the voltage causes a (small) change $\Delta v = e\,\Delta V / mv$ in speed, leading to a phase shift $\Delta\varphi = (2\pi s m/h)(e\,\Delta V / mv) = e\,\Delta V\,t_o/\hbar$. (This result is not altered by the fact that the energy $eV$ is shared by the electron and hole.) Thus, the modulation of the conductance with changing $V$ is due to ordinary electric forces; for that reason, it is not an AB effect.

## 5 Magnetic Analogue Observed?

A magnetic moment $\boldsymbol{\mu}$ in a uniform magnetic field $\mathbf{B}$ has magnetic potential energy $U = -\boldsymbol{\mu}\cdot\mathbf{B}$. Specifically, for a neutron in a uniform magnetic field, there is a shift in energy by +/- $\mu B$, corresponding to two orientations of the spin-1/2 angular momentum, compared with a neutron in a field-free region. Making use of this difference in energy, one might hope to observe an interference effect, analogous to the electrostatic AB proposal, in which the difference in energy along two interfering paths introduces a phase shift in the absence of any forces acting on the particle. Indeed, observation claims exist for this effect. Yet, as has been demonstrated above, conservation of energy requires that the energy change for the particle be balanced by an equal and opposite energy change for the apparatus; as interference takes place in the configuration space of the entire system, the resulting phase shifts cancel. An interference effect has been observed, yes, but it has been misinterpreted.

Referring back to Fig. 1, let us replace one of the cylinders with a solenoid, in which a current is turned on for a time t and turned off again, all while the neutron is safely within the interference apparatus. Because the neutron is at all times in a uniform field, there is no net magnetic force on it. Nevertheless, its energy on this path differs from its energy on the other path by an amount of magnitude $\mu B$ for time t, which would seem to generate a phase shift $\varphi = \mu B t/\hbar$. A phase difference of this magnitude has in fact been confirmed in observations of neutron interference [7,8]. How does it arise, if not from the energy difference?

The correct explanation is as follows. A magnetic moment $\mu$ along the z axis produces an azimuthal vector potential, at distance r in direction θ, given by

$A = (\mu_o/4\pi)\,\mu\,\sin\theta/r^2$.

If the z axis is also the axis of a solenoid, this vector acts on the moving charges that constitute the current in the wires, generating an azimuthal phase gradient (magnetic momentum). It is easy to demonstrate that the motion of the charges through this phase gradient generates a phase equal to μBt/ℏ in time t. What has been observed, therefore, is another example of the magnetic AB effect, not any analogue of the electrostatic version.

## 6  Concluding Remarks

The results presented here follow straightforwardly from standard quantum theory. Nevertheless, it is surprising that in what appears to be a single-particle interference experiment, one may have to take account of the wavefunction of the entire system, including the apparatus. (No doubt, that explains the incredulity with which this manuscript and earlier versions have been met by most physicists and philosophers of science with whom it has been shared.) A particle can travel two paths to a bank of detectors; is the pattern of detection not determined by the relative phases *of the particle* on those paths? No, not if the particle path itself sufficiently affects phases in a larger system. We must take seriously that quantum interference takes place in the many-dimensional configuration space of the entire system, even a macroscopic one.

## ACKNOWLEDGMENTS

The author thanks Kevin Martin for valuable comments and discussions, Mark Semon for pointing out Aharonov and Bohm's 1961 paper, Gordon Fleming for other useful literature references and for encouragement, and Sharon Bertsch for technical assistance in preparing the manuscript. Special thanks go to the referees at the *International Journal of Theoretical Physics* who were willing to set aside their initial skepticism long enough to give the paper a thorough evaluation.

1. Aharonov, Y., Bohm, D.: Phys. Rev. **115**, 485-489 (1959)
2. Chambers, R. G.: Phys. Rev. Lett. **5**, 3-5 (1960)
3. Tonomura, A., Osakabe, N., Matsuda, T., Kawasaki, T.: Phys. Rev. Lett. **56**, 792-795 (1986)
4. Sakurai, J. J.: *Modern Quantum Mechanics (Revised Edition)*, p. 124. (Addison Wesley, New York, 1993).
5. Aharonov, Y., Bohm, D.: Phys. Rev. **123**, 1511-1524 (1961)
6. Van Oudenaarden, A., Devoret, M. H., Nazarov, Yu. V., Mooij, J. E.: Nature **391**, 768-770 (1998)
7. van der Zouw, G., Weber, M., Felber, J., Gähler, R., Geltenbort, P., Zeilinger, A.: Nuclear Instruments and Methods in Physics Research A **440**, 568-574 (2000)
8. Werner, S. A. and Klein, A. G.: J. Phys. A: Math. Theor. **43**, 354006+33 (2010)

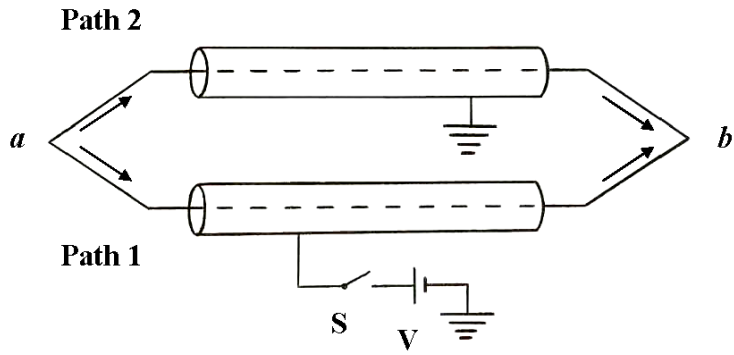

**Fig. 1** Aharonov and Bohm's original idealized experiment, showing two interfering paths of an electron from *a* to *b*. The electric potential along path 1 is changed from 0 to V and back to 0 while the wavepacket is within the cylinders.

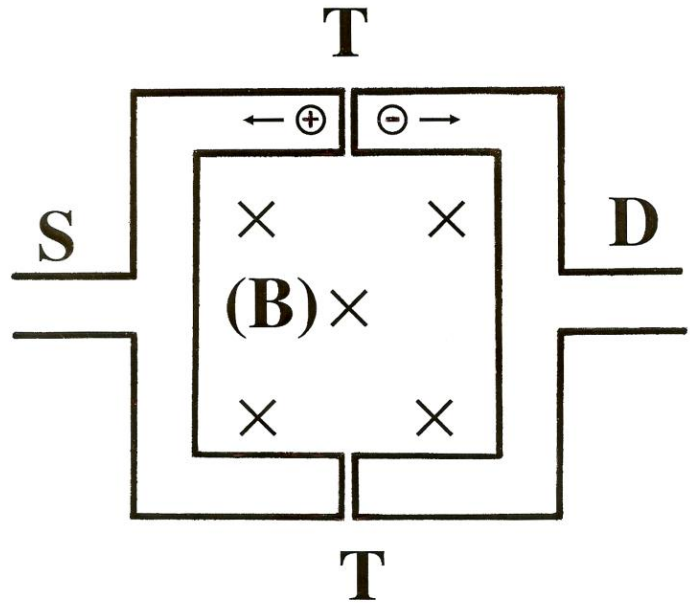

**Fig. 2** Interference in a mesoscopic metal ring. Conductance is measured between source (S) and drain (D). An electron-hole pair formed at one tunnel junction (T) may recombine at the other junction, or the particles may proceed to the drain and source.